\documentclass[,final]{aipproc}
\layoutstyle{6x9}
\usepackage{amsmath,amssymb}
\usepackage{mathrsfs}
\usepackage{bm}

\newcommand{\beq}{\begin{equation}}
\newcommand{\eeq}{\end{equation}}
\newcommand{\ba}{\begin{array}}
\newcommand{\ea}{\end{array}}
\newcommand{\bea}{\begin{eqnarray}}
\newcommand{\eea}{\end{eqnarray}}

\begin{document}
\title{Quark confinement 
due to non-Abelian magnetic monopoles 
in SU(3) Yang-Mills theory}


\author{Kei-Ichi Kondo}{
  address={Department of Physics,  Graduate School of Science, 
Chiba University, Chiba 263-8522, Japan
  },
}
\author{Akihiro Shibata}{
  address={Computing Research Center, High Energy Accelerator Research Organization,  Tsukuba  305-0801, Japan
  }
}
\author{Toru Shinohara}{
  address={Department of Physics,  Graduate School of Science, 
Chiba University, Chiba 263-8522, Japan
  }
}
\author{Seikou Kato}{
  address={Fukui National College of Technology, Sabae 916-8507, Japan
  }
}

\begin{abstract}
We present recent results on quark confinement:  in SU(3) Yang-Mills theory,  confinement of fundamental quarks is obtained due to the dual Meissner effect originated from non-Abelian magnetic monopoles defined in a gauge-invariant way, which is distinct from the well-known  Abelian projection scenario.  This is achieved by using  a non-Abelian Stokes theorem for the Wilson loop operator and a new reformulation of the Yang-Mills theory.   

\end{abstract}

\maketitle

\noindent
{\bf\emph{Introduction. --}}
The dual superconductor picture proposed long ago  \cite{dualsuper}  is believed to be a promising mechanics for quark confinement. 
For this mechanism to work, however, magnetic monopoles and their condensation are indispensable to cause the dual Meissner effect leading to the linear potential between quark and antiquark, namely,  
  area law of the Wilson loop average.  
The Abelian projection method proposed by 't Hooft \cite{tHooft81} can be used to introduce such magnetic monopoles into the pure Yang-Mills theory even without matter fields. 
Indeed, numerical evidences supporting the dual superconductor picture resulting from such magnetic monopoles have been accumulated since 1990  in pure SU(2) Yang-Mills theory \cite{SY90,SNW94,AS99}.
However, {\it the Abelian projection method explicitly breaks both the local gauge symmetry and the global color symmetry} by partial gauge  fixing from an original non-Abelian gauge group $G=SU(N)$ to the maximal torus subgroup, $H=U(1)^{N-1}$. 
Moreover, the Abelian dominance \cite{SY90} and  magnetic monopole dominance \cite{SNW94} were observed only in a special class of gauges, e.g., the maximally Abelian (MA) gauge and Laplacian Abelian (LA) gauge, realizing the idea of  Abelian projection.

For $G=SU(2)$, we have already succeeded to settle the issue of  gauge (in)dependence by {\it introducing  a gauge-invariant magnetic monopole in a gauge independent way}, based on another method: a non-Abelian Stokes theorem for the Wilson loop operator \cite{DP89,Kondo98b} and a new reformulation of Yang-Mills theory rewritten in terms of new field variables \cite{KMS06,KMS05,Kondo06} and \cite{KKMSSI05,IKKMSS06,SKKMSI07}, elaborating the technique proposed by Cho \cite{Cho80} and Duan and Ge \cite{DG79} independently, and later readdressed by Faddeev and Niemi \cite{FN99}.

For $G=SU(N)$, $N \ge 3$, there are no inevitable reasons why degrees of freedom associated with the maximal torus subgroup should be most dominant for quark confinement.
In this case, the problem is not settled yet. 
In this talk, 
we give a theoretical framework for describing {\it non-Abelian  dual superconductivity} in $D$-dimensional $SU(N)$ Yang-Mills theory, which should be compared with the conventional Abelian $U(1)^{N-1}$ dual superconductivity in $SU(N)$ Yang-Mills theory, hypothesized by  Abelian projection. 
We demonstrate that {\it an effective low-energy description for quarks in the fundamental representation} (abbreviated to rep. hereafter) {\it can be given by a set of non-Abelian restricted field variables} and that {\it non-Abelian $U(N-1)$ magnetic monopoles} in the sense of Goddard--Nuyts--Olive--Weinberg \cite{nAmm} {\it are the most dominant topological configurations for quark confinement} as conjectured in \cite{KT99,Kondo99Lattice99}.
This is the non-Abelian dual superconductor scenario for quark confinement for $SU(3)$ Yang-Mills theory proposed in \cite{KSSK10}.


\vspace*{1ex}
\noindent
{\bf\emph{Reformulation of Yang-Mills theory using new variables --}}
By using new variables, we have reformulated the SU(3) Yang-Mills theory in the continuum \cite{KSM08} and on a lattice \cite{lattice-f}. 
 For  $SU(3)$, there exist two possible options: maximal one with the maximal stability subgroup $\tilde H=U(1)^2$ \cite{Cho80c,FN99a} and the minimal one with the maximal stability subgroup $\tilde H=U(2)$ \cite{KSM08}. 
 The minimal option we have found in \cite{KSM08} is a new formulation. 
In our reformulation,  all the new variables are obtained by the change of variables from the original variable, once the color field $\bm{n}$ is determined by solving the reduction condition  for a given set of the original variables. 
In the continuum, for the change of variables  from   $\mathscr{A}_\mu$ to 
$\mathscr{C}_\mu$, $\mathscr{X}_\mu$ and $\bm{n}$: 
\begin{equation}
 \mathscr{A}_\mu^A  \Longrightarrow (\bm{n}^\beta, \mathscr{C}_\nu^k,  \mathscr{X}_\nu^b)  
 ,
\end{equation} 
the reduction condition is given by
\begin{equation}
\bm\chi[\mathscr{A},\bm{n}]
 :=[ \bm{n},  D^\mu[\mathscr{A}]D_\mu[\mathscr{A}]\bm{n} ]
 = 0 
  .
\label{eq:diff-red}
\end{equation}

\begin{figure}[h]
\includegraphics[height=6.0cm,width=10.0cm]{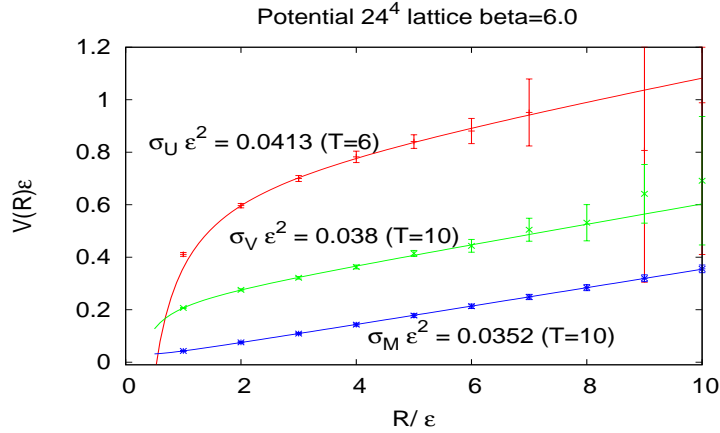}
\vspace{-0.6cm}
\caption{$SU(3)$ quark-antiquark potential: (from top to bottom) 
 full potential (red) $V_f(r)$,  restricted part (green) $V_a(r)$  and  magnetic--monopole part  (blue) $V_m(r)$  at $\beta=6.0$ on $24^4$ ($\epsilon$: lattice spacing).}
\label{fig:quark-potential}
\end{figure}

\vspace*{1ex}
\noindent
{\bf\emph{Numerical simulations on a lattice --}}
On a four-dimensional Euclidean lattice, the gauge field configurations (link variables)  $\{ U_{x,\mu} \}$ are generated by using the standard Wilson action and pseudo heat-bath method.  
For a given $\{ U_{x,\mu} \}$,  color field $\{ \bm{n}_{x} \}$ are determined by imposing a lattice version of  reduction condition.  Then new variables are introduced by using the lattice version of  change of variables \cite{lattice-f}.

In Fig. \ref{fig:quark-potential}, we give the result \cite{KSSK10}: the full $SU(3)$ quark-antiquark potential  $V(r)$ obtained from the Wilson loop average $\langle W_C[\mathscr{A}] \rangle$, the restricted part $V_a(r)$ obtained from the   Wilson loop average $\langle W_C[\mathscr{V}] \rangle$ of the restricted variable $\mathscr{V}:=\mathscr{A}-\mathscr{X}$, and  magnetic--monopole part  $V_m(r)$ obtained  from    $\langle e^{  ig_{\rm YM} (k, \Xi_{\Sigma})  }  \rangle$ following from the non-Abelian Stokes theorem. They are gauge invariant by construction.  These results exhibit infrared $\mathscr{V}$ dominance in the string tension (85--90\%) and  non-Abelian $U(2)$ magnetic monopole dominance in the string tension  (75\%) in the gauge independent way.

\begin{figure}[bt]
\centering
\includegraphics[height=4.5cm,]{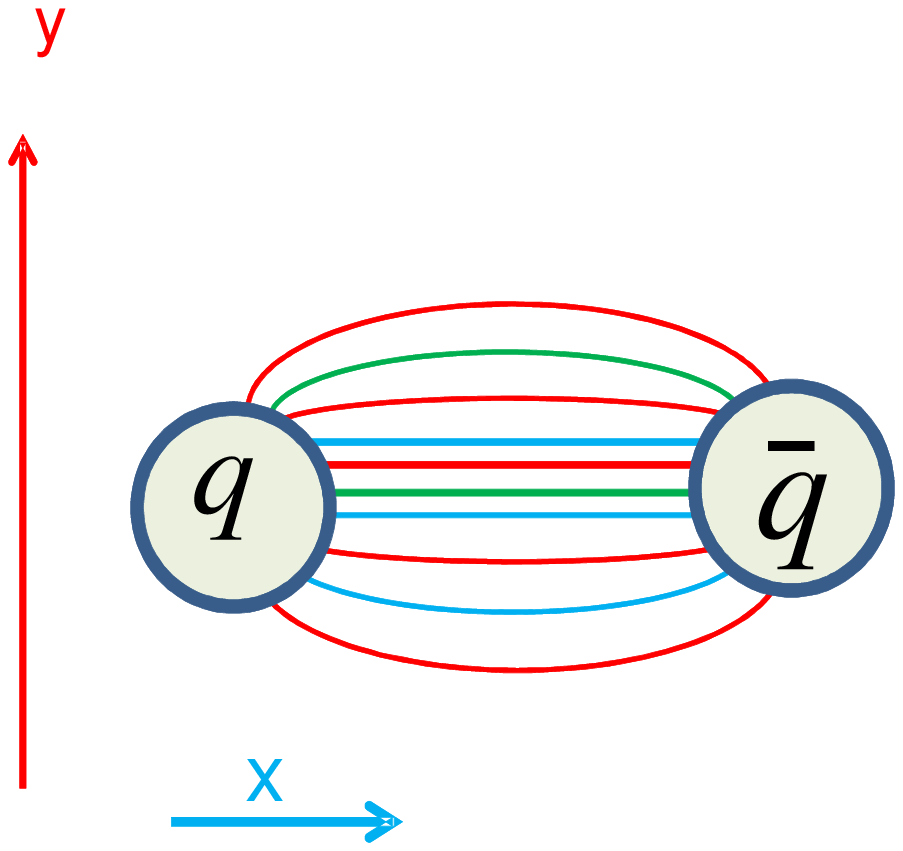} 
\includegraphics[origin=c,height=4.5cm,angle=270]{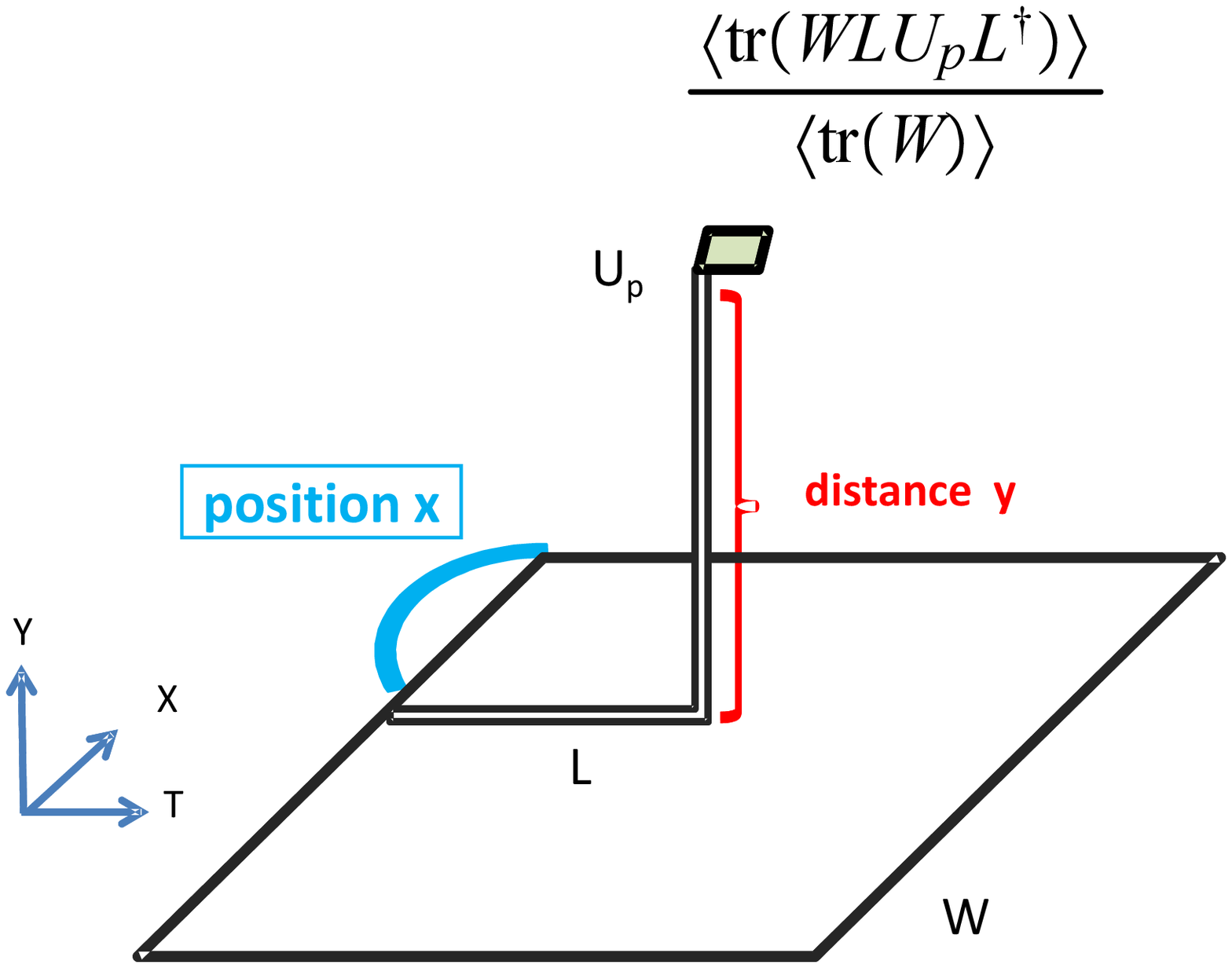}
\caption{
(Left panel) The setup for measuring the color flux produced by a
quark--antiquark pair. (Right panel) The connected correlator between a plaquette $P$
and the Wilson loop $W$.
}%
\label{Fig:Operator}%
\end{figure}

\begin{figure}[bt]
\centering
\includegraphics[height=7cm,angle=270]{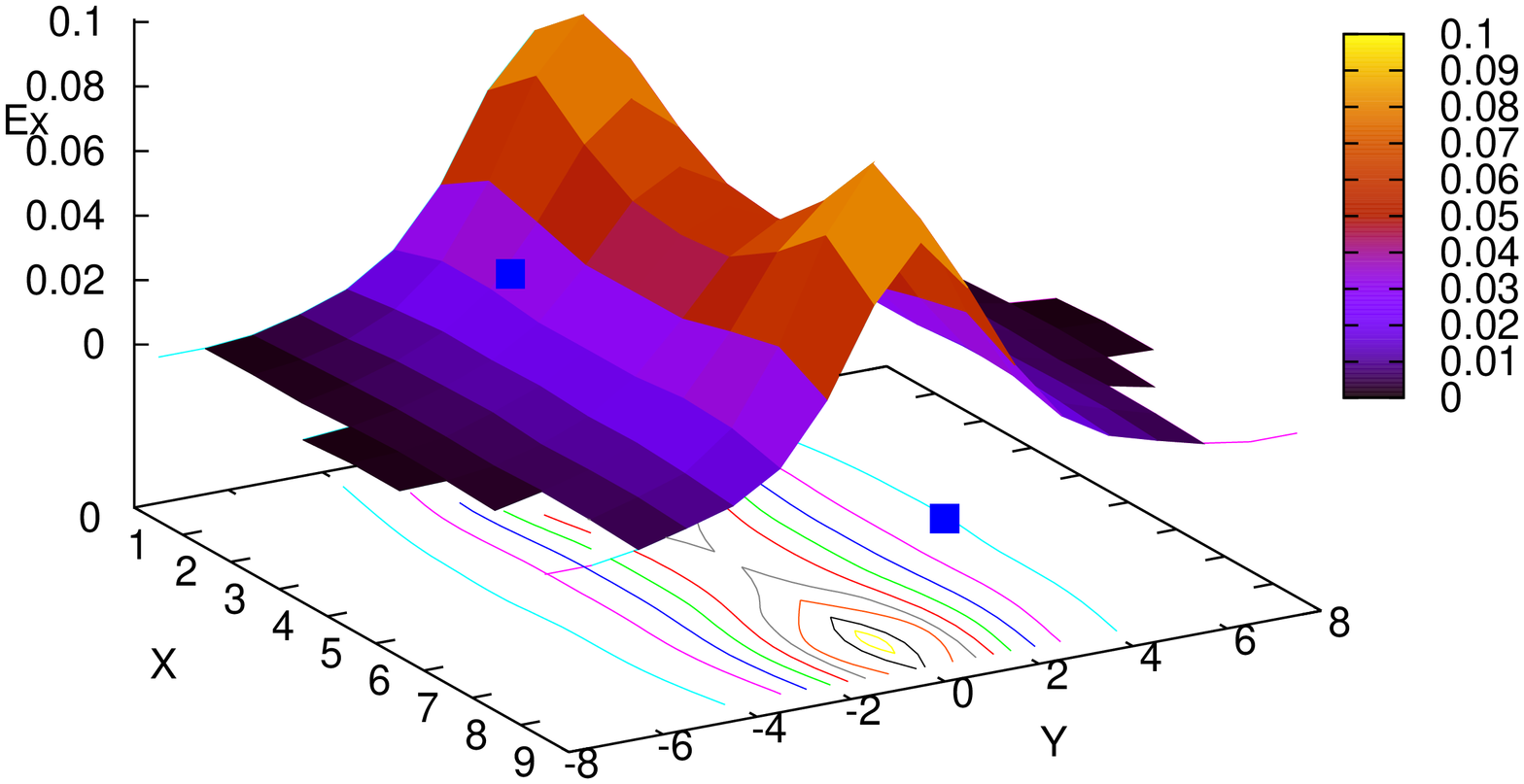} 
\includegraphics[height=7cm,angle=270]{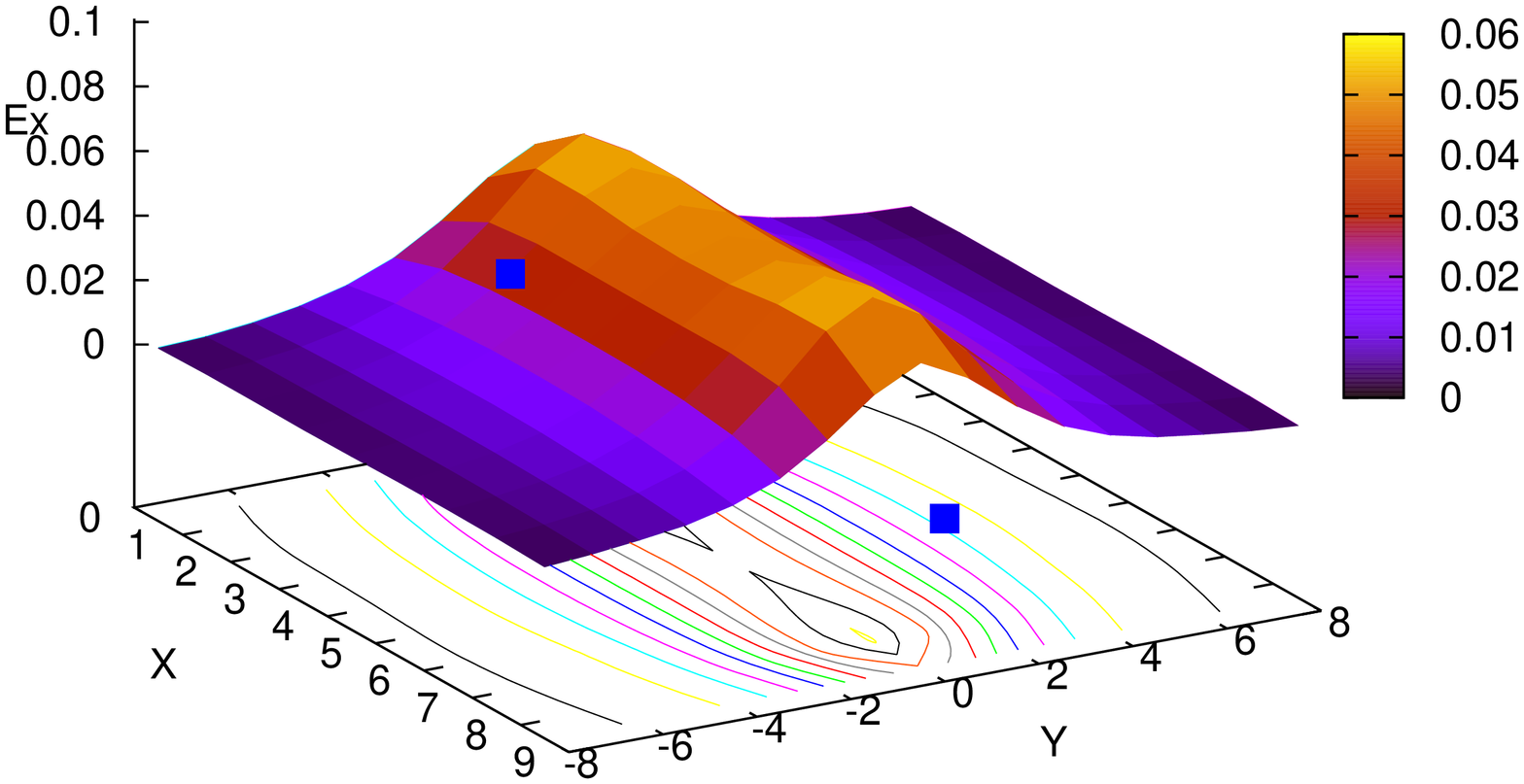}
\caption{
Measurement  of chromo-electric flux in the SU(3) Yang-Mills theory.  (Left)   $E_{x}$ from the original field $\mathscr{A}$. (Right) \ $E_{x}$   from the restricted  field $\mathscr{V}$.
}%
\label{fig:fluxtube}%
\end{figure}

In Fig.\ref{Fig:Operator}, we give  the color flux produced by a quark-antiquark pair obtained in \cite{Shibata12}. 
In order to explore the color flux in the
gauge invariant way, we use the connected correlator $\rho_{W}$ of the Wilson line
  (see the right panel\ of Fig.\ref{Fig:Operator}):
\begin{equation}
\rho_{W}:=\frac{\left\langle \mathrm{tr}\left(  U_{P}L^{\dag}WL\right)
\right\rangle }{\left\langle \mathrm{tr}\left(  W\right)  \right\rangle
}-\frac{1}{N}\frac{\left\langle \mathrm{tr}\left(  U_{P}\right)
\mathrm{tr}\left(  W\right)  \right\rangle }{\left\langle \mathrm{tr}\left(
W\right)  \right\rangle }.
 \label{eq:Op}%
\end{equation}
In the naive continuum limit,   $\rho_{W}$ reduces to the field strength:
\begin{equation}
\rho_{W}\overset{\varepsilon\rightarrow0}{\simeq}g\epsilon^{2}\left\langle
\mathcal{F}_{\mu\nu}\right\rangle _{q\bar{q}}:= \left\langle
\mathrm{tr}\left(  g\epsilon^{2}\mathcal{F}_{\mu\nu}L^{\dag}WL\right)
\right\rangle /\left\langle \mathrm{tr}\left(  W\right)  \right\rangle
 +O(\epsilon^{4}) .
\end{equation}
Thus, the color filed strength produced by a $q\bar q$ pair is given by $\ F_{\mu\nu}=\sqrt{\frac{\beta
}{2N}}\rho_{W}$.

These are numerical evidences supporting  the {``non-Abelian'' dual superconductivity due to non-Abelian  magnetic monopoles   as a mechanism for quark confinement in SU(3) Yang-Mills theory}.

\vspace*{1ex}
\noindent
{\bf\emph{Summary. --}}
We have shown for the $SU(N)$ Yang-Mills theory in $D$-dimensions: 

\begin{enumerate}
\item[(a)] 
We have defined  {a gauge-invariant magnetic monopole $k$   inherent in the Wilson loop operator} by using a non-Abelian Stokes theorem for the Wilson loop operator, even in $SU(N)$ Yang-Mills theory without adjoint scalar fields. 
The $SU(N)$ Wilson loop operator can be rewritten \cite{Kondo08} in terms of a pair of the gauge-invariant magnetic-monopole current  $k$ ($(D-3)$-form) and the associated geometric object defined from the Wilson surface $\Sigma$ bounding the Wilson loop $C$, and another pair of an electric current $j$ (one-form  independently of $D$) and the associated topological object, due to a non-Abelian Stokes theorem for the Wilson loop operator \cite{Kondo08}.

For quarks in the fundamental representation, the stability group is given by $\tilde H=U(N)$ for $G=SU(N)$.
\\
$\bullet$ G=SU(2) Abelian magnetic monopole SU(2)/U(1)
\\
$\bullet$ G=SU(3) non-Abelian magnetic monopole SU(3)/U(2)


\item[(b)]  
We have constructed a new reformulation  \cite{KSM08} of the $SU(N)$ Yang-Mills theory  in terms of new field variables obtained by change of variables from the original Yang-Mills gauge field $\mathscr{A}_\mu^A(x)$, so that it gives an optimal description of the non-Abelian magnetic monopole defined from the $SU(N)$ Wilson loop operator in the fundamental rep. of quarks.
The reformulation allows options discriminated by the maximal stability group $\tilde{H}$ of the gauge group $G$.  

For $G=SU(3)$, two options are possible: 

\noindent
$\bullet$ The minimal option $\tilde{H}=U(2)$ gives an optimized description of quark confinement through the Wilson loop in the fundamental representation.

\noindent
$\bullet$ The maximal option, $\tilde{H}=H=U(1) \times U(1)$, the new theory reduces to a manifestly gauge-independent reformulation of the conventional Abelian projection in the maximal Abelian gauge.

The idea of using new variables is originally due to Cho \cite{Cho80c}   and Faddeev and Niemi \cite{FN99a}, where $N-1$ color fields $\bm{n}_{(j)}$ ($j=1,...,N-1$) are introduced. 
However, our reformulation in the minimal option is new for $SU(N), N \ge 3$: we introduce  {only a single color field $\bm{n}$ for any $N$}, which is enough for reformulating the quantum Yang-Mills theory to describe confinement of the  {fundamental quark}.

\item[(c)]  
We have constructed a lattice version \cite{lattice-f} of the reformulation of the $SU(N)$ Yang-Mills theory and performed numerical simulations for the $SU(3)$ case on a lattice,
Numerical simulations of the lattice $SU(3)$ Yang-Mills theory give numerical evidences that the restricted field variables become dominant in the infrared for the string tension and correlation functions ({\it infrared dominance of the restricted non-Abelian variables}) and that the $U(2)$ magnetic monopole gives a dominant contribution to the string tension obtained from  $SU(3)$ Wilson loop average ({\it non-Abelian magnetic monopole dominance} for for quark confinement (in the string tension)). 
This should be compared with the infrared Abelian dominance and magnetic monopole dominance in MA gauge.

\item[(d)]  
We have shown the numerical evidence of the  {dual Meissner effect caused by non-Abelian magnetic monopoles} in $SU(3)$ Yang-Mills theory:  the tube-shaped flux of the chromo-electric field originating from the restricted field including the non-Abelian magnetic monopoles. 

\end{enumerate}

To confirm the  {non-Abelian dual superconductivity  picture proposed \cite{KSSK10} for $SU(3)$ Yang-Mills theory}, we plan to do further checks, e.g., determination of the type of dual superconductor, measurement of the penetrating depth, induced magnetic current around color flux due to magnetic monopole condensations, and so on.

\vspace*{1ex}
\noindent
{\bf\emph{Acknowledgments. --}}
This work is supported by Grant-in-Aid for Scientific Research (C) 24540252 from Japan Society for the Promotion Science (JSPS), and also in part by JSPS Grant-in-Aid for Scientific Research (S) 22224003. 
The numerical calculations are supported by the Large Scale Simulation Program No.09/10-19 (FY2009-2010) and No.10-13 (FY2010) of High Energy Accelerator Research Organization (KEK).

\bibliographystyle{aipproc}   

\end{document}